# Projection-based measurement and identification

C. Jailin[1,2], A. Buljac[1], A. Bouterf[1], F. Hild[1] and S. Roux[1]

[1] LMT (ENS Paris-Saclay/CNRS/Univ.Paris-Saclay), 61 avenue du Pdt. Wilson, 94235 Cachan (FRANCE)

[2] Safran Aircraft Engines - SAE, Rond-Point René Ravaud, Réau, 77550 Moissy-Cramayel (FRANCE)

**ABSTRACT**

A recently developed Projection-based Digital Image Correlation (P-DVC) method is here extended to 4D (space and time) displacement field measurement and mechanical identification based on a single radiograph per loading step instead of volumes as in standard DVC methods. Two levels of data reductions are exploited, namely, reduction of the data acquisition (and time) by a factor of 1000 and reduction of the solution space by exploiting model reduction techniques. The analysis of a complete tensile elastoplastic test composed of 127 loading steps performed in 6 minutes is presented. The 4D displacement field as well as the elastoplastic constitutive law are identified.

**INTRODUCTION**

Identification and validation of increasingly complex mechanical models is a major concern in experimental solid mechanics. The recent developments of computed tomography coupled with *in-situ* tests provide extremely rich and non-destructive analyses [1]. In the latter cases, the sample was imaged inside a tomograph, either with interrupted mechanical load or with a continuously evolving loading and on-the-fly acquisitions (as ultra-fast X-ray synchrotron tomography, namely, 20 Hz full scan acquisition for the study of crack propagation [2]). Visualization of fast transformations, crack openings, or unsteady behavior become accessible. Combined with full-field measurements, *in-situ* tests offer a quantitative basis for identifying a broad range of mechanical behavior.

A now common method to quantitatively measure kinematic data from the reconstructed images is Digital Image Correlation (DIC) in 2D and its 3D extension, Digital Volume Correlation (DVC) [3]. The latter aims at capturing the way a solid deforms between two states from the analysis of the corresponding 3D images. The measured displacement field is then used to calibrate model parameters from inverse procedures (*e.g.*, finite element model updating, virtual fields method). The more numerous the acquisitions (in space and time), the more accurate and sensitive the identification procedure. Those 4D (space-time) analyses (*e.g.* [4]) always consist of a sequence of three successive inverse problems: (i) volume reconstructions, (ii) kinematic measurements from Digital Volume Correlation (DVC), and (iii) constitutive law identification (see Figure 1). These approaches, often limited to few scans, are hence extremely dense in space measurements but suffer from sparse time sampling.

A short-cut to the previously described three steps sequence, which is called Projection-based DVC (P-DVC) [5-7], evaluates the full 4D kinematics directly from few selected projections and one single reference volume. The number of radiographs needed for tracking the 3D space plus time changes of the test is thereby reduced from 500-1000 down to a single one per time step. With this procedure, the sample is continuously loaded, continuously rotated and regularly imaged by 2D X-ray projections.

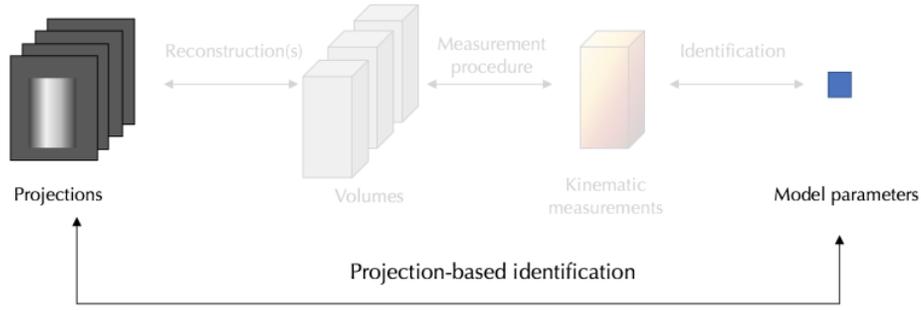

Figure 1: Dataflow in tomography measurement composed of 3 inverse problems and its short-cut called Projection-based DVC.

**BACKGROUND**

First, tomography consists in reconstructing a 3D volume $f(x)$ for all space locations $x$ from sets of radiographs $p(r,\theta)$ acquired on the detector $r$ at angle $\theta$. Among many tomographic reconstruction algorithms, algebraic schemes are based on the minimization of the following functional with respect to the image voxel intensity

$$\Gamma_{\text{ART}} = \sum_{r,\theta} \|\Pi_\theta[f(x)] - p(r,\theta)\|^2$$

where $\Pi_\theta$ is the projection operator in the $\theta$ direction. Digital Volume Correlation (DVC) [3-4] is a full field measurement technique for the 4D displacement field that relates a series of 3D images, namely, one volume for the reference state $f(x)$ and few images for the deformed states $g(x,t)$ indexed by time $t$. The DVC procedure (written here with the Eulerian transformation to unify notations, considering the next functional) corresponds to the minimization of the quadratic difference between the reference image corrected by the measured displacement $u(x,t)$ and the images of the deformed state

$$\Gamma_{\text{DVC}} = \sum_{x,t} \|f(x + u(x,t)) - g(x,t)\|^2$$

A kinematic regularization of the displacement field is introduced in global DVC [8] for which the displacement field is expressed on a reduced basis, composed of a set of $N_T$ time functions $\sigma(t)$ and $N_s$ space fields $\Phi(x)$ such that

$$u(x,t) = \sum_{i=1}^{N_T}\sum_{j=1}^{N_s} u_{ij}\sigma_i(t)\Phi_j(x)$$

where $u_{ij}$ are the kinematic unknowns. A general framework for the kinematic bases well suited to mechanical modeling is the framework used in the finite element method. The displacement field is obtained from the minimization of the functional with respect to the space-time degrees of freedom $u_{ij}$ (i.e., the nodal displacements). $N_s N_T$ defines the number of degrees of freedom.

The proposed approach to fast 4D (space and time) measurements is called Projection-based Digital Volume Correlation (P-DVC). The registration consists in minimizing the sum of squared differences between $N_t$ 2D projections $p(r,\theta(t))$ of the deformed configuration, acquired at different times $t$ or angles $\theta(t)$, and loading steps

$$\Gamma_{\text{P-DVC}} = \sum_{r,t} \left\|\Pi_{\theta(t)}[f(x + u(x,t))] - p(r,\theta(t))\right\|^2$$

The procedure makes use of the 3D reference image, $f(x)$, which is reconstructed using classical means (e.g. before the experiment, in a static configuration). As in 3D analyses, a key quantity to validate the procedure is the residual (error) field defined, in P-DVC, as the difference between the projection of the corrected volume and the original projections: $\Pi_{\theta(t)}[f(x + u(x,t))] - p(r,\theta(t))$. Initially, it contains large errors essentially due to the motion. At convergence, these fields indicate what was not captured by the correction model such as incorrect kinematic model assumptions (erroneous choice of $\Phi(x)$ and $\sigma(t)$), noise and reconstruction artifacts. Last, the displacement field is here measured using successive mode identification (PGD measurements) using a greedy algorithm [9-10].

**ANALYSIS**

The present PGD – P-DVC procedure is tested with the analysis of a tensile test on a dog-bone sample made of nodular graphite cast iron until failure. The sample was scanned in the LMT lab-CT (NIS X50+). The test is composed of two main parts (see Figure 2):

- acquisition of a reference static volume (with a small preload) that took 22 minutes,

- continuous rotation of the sample with 50 projections per full rotation at a rate of one projection every 2 s. One hundred twenty-seven projections were acquired during 2.5 full rotations in 6 minutes. The first full rotation (*i.e.*, 50 time steps or 100 s) was performed at constant load and was used to quantify the uncertainty. The remaining rotation (starting after 100 s) was carried out with a continuous load change (from 250 to 750 N), as shown in Figure 4, controlled at a constant stroke velocity of 2 µm/s.

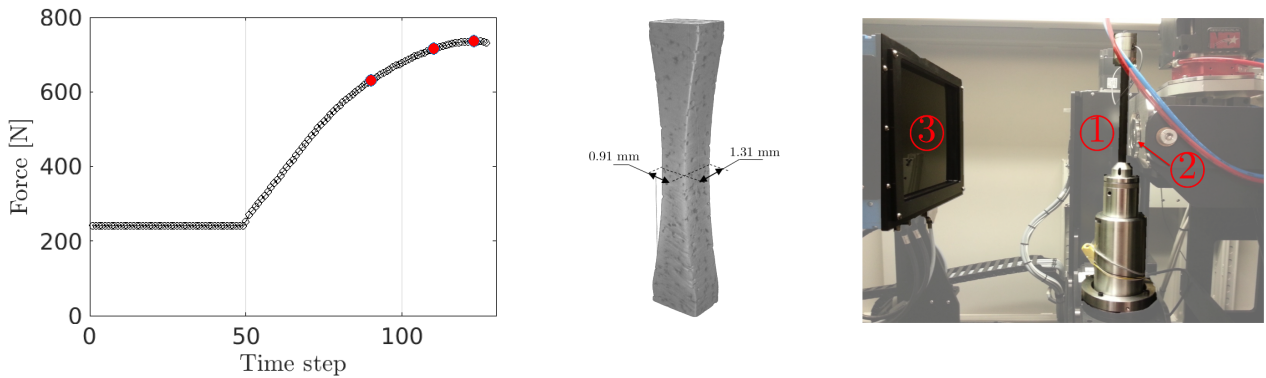

Figure 2: (left) Loading for the 127 acquisitions, performed in 6 minutes. The projections shown in Figure 3 are the red dots. (center) Scanned dog-bone geometry. (right) I*n-situ* testing device in the tomograph.

After a rigid body motion correction, the full kinematics is identified. The space is regularized by a kinematics (derived from beam theory) driven by 15 rigid sections and trilinear interpolations (*i.e.*, $N_s = 15 \times 6$ degrees of freedom). The time is regularized by $N_T = 7$ time functions (force measurements, polynomials, sin/cosine). In total, 630 space-time degrees of freedom are finally measured.

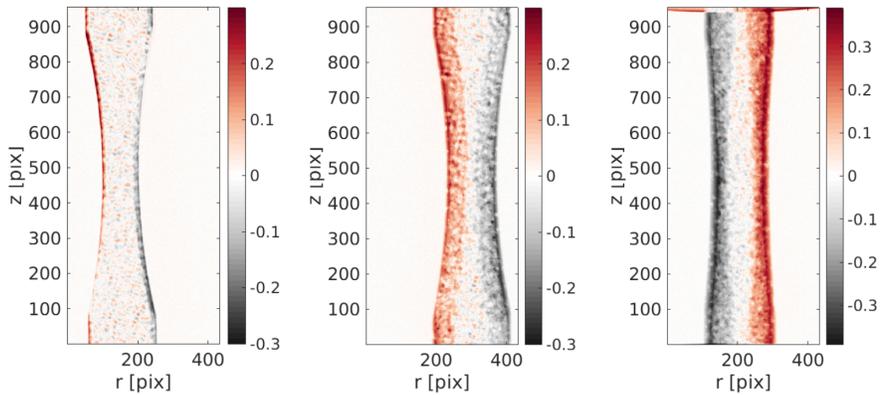

Figure 3: Three selected initial projected residual fields (see Figure 2). The high positive and negative values are the signature of motions.

The change of the 127 projected residual fields are shown for 3 selected angles (*i.e.* acquired at different rotations and loads, see red dots of Figure 2) and are located at the end of the loading sequence, hence correspond to some of the largest strains. The initial and final residuals are shown in Figures 3 and 4. A large part of the motion patterns has been erased. Even black parts are appearing in the right image (just before failure) and are due to strain localization. The signal to noise ratio of the 127 residual fields increases from 9.9 dB to 26.6 dB.

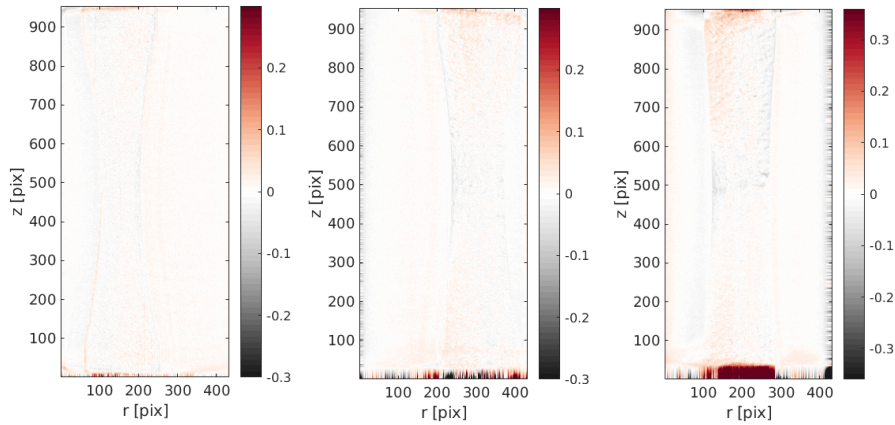

Figure 4: Three selected projected residual fields, after the kinematic correction. It can be seen that almost all motion patterns have been erased.

The measured vertical displacement field for each longitudinal section and time is shown in Figure 5(left). From the measured force at each time step, a very simple 1D behavior of the beam is identified. The results of the stress-strain identification for all measurements is shown in Figure 5(right).

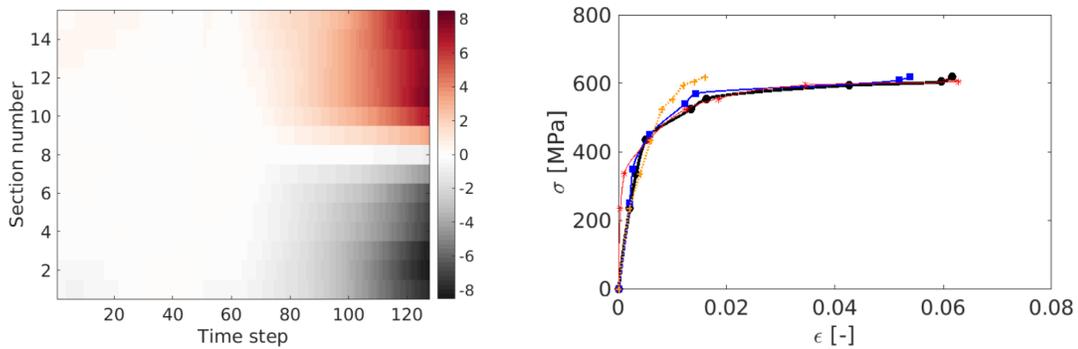

Figure 5: (left) measured vertical displacement field for each 15 sections of the beam as a function of time expressed in voxels (1 voxel ↔ 10.7 μm). (right) Identified stress-strain behavior.

**CONCLUSION**

The full space time kinematics of the 6 minute experiment was captured with an extension of DVC called P-DVC that uses a model order reduction technique (PGD). Based on highly regularized fields relying on the slender sample geometry as well as a dense sampling in time, this method measures displacement fields from single projections at each time (or load) step of the experiment instead of reconstructed volumes in standard DVC methods. The procedure was tested with an *in-situ* tensile test on nodular graphite cast iron composed of 127 radiographs with continuous load changes and rotations of the sample until failure. The experiment was carried out in a lab tomograph with an X-ray cone beam source.

The major advantage of PDVC is the important time sampling (and hence temporal resolution). The entire experiment was carried out in 300 s, which is more than two orders of magnitude faster than standard methods. This performance goes together with the benefit of having a continuous (*i.e.*, uninterrupted) loading so that load and rotation can be varied simultaneously. This method could be coupled with 3D DVC analysis and thus benefits both from the space (DVC) and time (P-DVC) huge resolution.

**ACKNOWLEDGEMENTS**

This work has benefited from the support of the French "Agence Nationale de la Recherche" through the "Investissements d'avenir" Program under the reference ANR-10-EQPX-37 MATMECA, and ANR-14-CE07-0034-02 COMINSIDE.

**REFERENCES**

[1] E. Maire, J. Y. Buffière, L. Salvo, J. J. Blandin, W. Ludwig, J. M. Létang On the application of X-ray microtomography in the field of materials science. Advanced Engineering Materials, 3(8):539–546, 2001.

[2] E. Maire, C. Le Bourlot, J. Adrien, A. Mortensen, and R. Mokso. 20 Hz X-raytomography during an in situ tensile test. International Journal of Fracture, 200(1):3–12, 2016.

[3] B.K. Bay, T.S. Smith, D.P. Fyhrie, and M. Saad. Digital volume correlation: three-dimensional strain mapping using X-ray tomography. Experimental mechanics, 39(3):217–226, 1999.

[4] F. Hild, A. Bouterf, L. Chamoin, H. Leclerc, F. Mathieu, J. Neggers, F. Pled, Z. Tomičević & S. Roux. Toward 4D Mechanical Correlation. Advanced Modeling and Simulation in Engineering Sciences, 3:17, 2016.

[5] T. Taillandier-Thomas, S. Roux & F. Hild. A soft route toward 4D tomography. Physical Review Letters, 117:025501, 2016.

[6] C. Jailin, A. Bouterf, M. Poncelet & S. Roux. In situ µ-CT-scan mechanical tests: Fast 4D mechanical identification. Experimental Mechanics, 57:1327–1340, 2017.

[7] H. Leclerc, S. Roux, and F. Hild. Projection savings in CT-based digital volume correlation. Experimental Mechanics, 55(1):275–287, 2015.

[8] S. Roux, F. Hild, P. Viot, and D. Bernard. Three-dimensional image correlation fromX-ray computed tomography of solid foam. Composites Part A: Applied science and manufacturing, 39(8):1253–1265, 2008.

[9] J.C. Passieux & J.N. Périé. High resolution digital image correlation using proper generalized decomposition: PGD-DIC. International Journal for Numerical Methods in Engineering, 92:531–550, 2012.

[10] L.A. Gomes Perini, J.C. Passieux & J.N. Périé. A Multigrid PGD-based Algorithm for Volumetric Displacement Fields Measurements. Strain, 50:355–367, 2014.